# Towards a Reconceptualisation of Cyber Risk: An Empirical and Ontological Study

Alessandro Oltramari, Bosch Research and Technology Center,
alessandro.oltramari@us.bosch.com

Alexander Kott, U.S. Army Research Laboratory,
alexander.kott1.civ@mail.mil

## Abstract

The prominence and use of the concept of cyber risk has been rising in recent years. This paper presents empirical investigations focused on two important and distinct groups within the broad community of cyber-defence professionals and researchers: (1) cyber practitioners and (2) developers of cyber ontologies. The key finding of this work is that the ways the concept of cyber risk is treated by practitioners of cybersecurity is largely inconsistent with definitions of cyber risk commonly offered in the literature. Contrary to commonly cited definitions of cyber risk, concepts such as the likelihood of an event and the extent of its impact are not used by cybersecurity practitioners. This is also the case for use of these concepts in the current generation of cybersecurity ontologies. Instead, terms and concepts reflective of the adversarial nature of cyber defence appear to take the most prominent roles. This research offers the first quantitative empirical evidence that rejection of traditional concepts of cyber risk by cybersecurity professionals is indeed observed in real-world practice.

## Introduction
The term 'cyber risk' refers to a variety of phenomena that damage or otherwise affect in an undesirable manner the information and information-related technology assets of a firm, an individual or a government institution. Common examples include identity theft, disclosure of sensitive information, and operations interruption. Many attempts have been made to define cyber risk. Some of these definitions, and their limitations, are discussed in later sections of this paper.
The prominence of cyber risk has been rising in the last few years, partly due to the growing recognition of such risks to industries and governments. Highly damaging—in terms of both money and reputation—cyberattacks are routinely and widely discussed in the media, which call into question the adequacy of defensive measures and insurance protection against the risks of such incidents.
The insurance industry has initiated significant efforts to research cyber risk and to develop insurance products for cyber risks. Such products are beginning to be seen as a necessity for doing business (Grande 2014). Biener, Eling, and Wirfs (2014) describe the state of the insurance industry with respect to cyber risk, note its rapid growth, and identify a number of problems with insurability of cyber risk. Modelling cyber risk and collecting empirical data for validation of such models are critical to resolving the insurability problems.

The United States government also pursues a cybersecurity strategy that involves monitoring and assessing cyber risks (Dempsey *et al.* 2011). For example, in 2010, the Office of Management and Budget directed federal agencies to use automated continuous monitoring and risk models (Executive Office of the President 2010). The Chief Information Office in the U.S. Department of Defense (DoD) recently released a document on its mobile device strategy that specifies continuous risk monitoring as part of the device management service (CIO 2012b), which was immediately followed by a document on cloud computing strategy that gives a similarly major role to risk monitoring (CIO 2012a).

Risk monitoring involves collection of data through automated feeds including network traffic information as well as host information from host-based agents; vulnerability information and patch status about hosts on the network; scan results from tools like Nessus; Transmission Control Protocol (TCP) netflow data; and Domain Name Server (DNS) trees, among others. These data undergo automated analysis in order to assess the risks. The assessment may include flagging especially egregious vulnerabilities and exposures or computing metrics that provide an overall characterisation of the network's risk level. In current practice, risk metrics are often simple sums or counts of vulnerabilities, missing patches, exploits, and impact of vulnerability exploitations.

There are important benefits in automated quantification of risk, for example assigning risk scores or other numerical measures to the network as a whole, its subsets, and even individual assets (Kott & Arnold 2013). This quantification opens doors to comprehensive risk-management decision-making that is, potentially, highly rigorous and insightful. Employees at multiple levels—from senior leaders to system administrators—will be aware of continually updated risk distribution over the network components and will use this awareness to prioritise application of resources to the most effective remedial actions. Quantification of risks is not only important to improve human decision processes, but it can contribute to rapid, automated, or semi-automated implementation of remediation plans.

Such significant potential benefits encourage funding of active efforts in cyber-risk research. For example, The U.S. Army Research Laboratory initiated a Cyber Collaborative Research Alliance program (Cyber CRA), a major part of which is focused on developing a comprehensive scientific approach to cyber risk (McDaniel, Rivera & Ananthram 2014). From the policy perspective, the National Institute of Standards and Technology (NIST) has been developing and publishing a steady stream of influential documents that, among other topics of cybersecurity, provide guidelines pertaining to cyber risk (http://www.nist.gov/cyberframework/).

This paper explores empirical investigations focused on two important and distinct groups within the broad community of cyber-defence professionals and researchers: (1) cyber practitioners and (2) developers of cyber ontologies. The key finding and main contribution of this work is that the ways the concept of cyber risk is treated by practitioners of cybersecurity is largely inconsistent with definitions of cyber risk commonly offered in the literature. Particularly salient is the observation that, contrary to commonly cited definitions of cyber risk, concepts such as the likelihood of an event and the extent of its impact are not used by cybersecurity practitioners. This is also the case for use of these concepts in the current generation of cybersecurity ontologies. Instead, terms and concepts reflective of the adversarial nature of cyber defence appear to take the most prominent roles.

Therefore, the authors argue that defender-centric models and their related definitions, ontologies, and risk-quantification approaches capture, at best, a subset of the essential dimensions of cyber risk; these models and definitions are, therefore, insufficient to deal with the complexity of risk in real-world scenarios. One might conjecture this has been long suspected by researchers of cyber risk. For example, Cox (2008) used theoretical arguments to reject conventional definitions of risk in application to adversarial contexts, and to warn against forcing practitioners to use conceptually meaningless definitions. His arguments, however, left open the possibility that the practitioners themselves might somehow adapt to the prevailing conceptualisation, and might find a way to use it effectively. To the authors' knowledge, the results presented in this paper constitute the first quantitative, empirical evidence that practitioners do, in fact, decline to use the prevailing conventional concepts of cyber risk. This paper makes the case for a conceptual and formal extension of the current notion of cyber risk, one in which the adversarial aspect of cyber decision-making is given its due centrality.

The remainder of the paper is organised as follows. The next three sections provide an overview of prior and ongoing work related to the conceptualisation of cyber risks from three perspectives: definitions of cyber risk; ontologies of cyber risk; and quantification of cyber risk. Then, the authors present methods and findings of two empirical studies. The first empirical study examines the ways in which cyber practitioners describe cyber risk. The second looks at how developers of computational ontologies conceptualise cyber risk. Finally, the authors offer a section of discussion and conclusions, which focuses on the need to give much greater weight to the adversarial perspective in defining, representing, and quantifying cyber risk.

## Definitions of Cyber Risk

The literature on cyber risk (including what some call 'IT risk') most commonly defines it in terms that combine likelihood of an undesirable event and a measure of the impact of the event.

For example, in the National Institute of Standards and Technology's (NIST) Special Publication 800-30, *Guide for conducting risk assessments*, risk is described as

> a function of the likelihood of a given threat-source's exercising a particular potential vulnerability, and the resulting impact of that adverse event on the organization. To determine the likelihood of a future adverse event, threats to an IT system must be analyzed in conjunction with the potential vulnerabilities and the controls in place for the IT system. (2012)

Similarly, in *Minimum security requirements for federal information and information systems*, NIST FIPS 2006, cyber risk is defined as the level of impact on organisational operations (including mission, functions, image, or reputation), organisational assets, or individuals resulting from the operation of an information system given the potential impact of a threat and the likelihood of that threat's occurring (NIST 2006).

The International Organization for Standardization and the International Electrotechnical Commission (ISO/IEC) (2008) definition of IT risk is also similar: the potential that a given threat will exploit vulnerabilities of an asset or group of assets and thereby cause harm to the organisation. It is measured in terms of a combination of the probability of occurrence of an event and its consequence.

National Security Telecommunications and Information Systems Security Instruction (NSTISSI) No. 1000 (NSTISSC 2000) also uses the notion of likelihood and defines risk as a combination of the likelihood that a threat will occur, the likelihood that a threat occurrence will result in an adverse impact, and the severity of the resulting impact.

National Information Assurance Training and Education Center (NIATEC) defines risk by listing several key aspects (NIATEC 2006):
- the loss potential that exists as the result of threat-vulnerability pairs;
- the uncertainty of loss expressed in terms of probability of such loss;

- the probability that a hostile entity will successfully exploit a particular telecommunication or Communications Security (COMSEC) system for intelligence purposes; its factors are threat and vulnerability;
- a combination of the likelihood that a threat shall occur, the likelihood that a threat occurrence shall result in an adverse impact, and the severity of the resulting adverse impact;
- the probability that a particular threat will exploit a particular vulnerability of the system.

The Open Web Application Security Project (OWASP) Testing Guide outlines a risk rating methodology (OWASP 2016) that starts with the standard risk model where Risk = Likelihood x Impact, and proceeds with recommendations for assigning specific numbers for likelihood and impact.

All of the above definitions rely largely on some combination of likelihood and the extent of the impact. Some add considerations of threat and vulnerability. However, there are also examples of a different approach. For instance, in Instruction No. 4009, the Committee on National Security Systems (CNSS) of the United States of America defines risk as the "possibility that a particular threat will adversely impact an information system by exploiting a particular vulnerability" (2010). This definition eschews any mention of the extent of the likelihood or the impact. The following section presents empirical observations that motivate the reluctance to include the commonly accepted notions of likelihood and impact into the definition of cyber risk.

Another vulnerability-centric definition of cyber risk can be found in NIST's Common Vulnerability Scoring System (CVSS), an open framework for computing the severity of software vulnerabilities. Vulnerability scores leverage on three metrics: (1) Base, (2) Temporal, and (3) Environmental. 'Base' includes the intrinsic properties of vulnerabilities, and represents the basic score—ranging from 0 to 10—upon which time-sensitive properties may occur (2), and measures related to user's environment (3) are computed. The resulting score indicates the severity of the vulnerability and, derivatively, the severity of the associated risk.

## Ontologies of Cyber Risk

An ontology provides a model of concepts in the domain of interest in a more elaborate, formal, and exhaustive manner than conventional definitions. Every science is concerned with distinct objects and strives to build rigorous models of the phenomena involving them (Bunge 1979). Accordingly, the objects of a science of cybersecurity correspond to the attributes of (and the relations between) a network of computer devices, security policies, and the tools and techniques of cyber attack and cyber defence (Kott 2014). Therefore, inasmuch as ontologies are formal models of a domain, building ontologies of the aforementioned attributes and relations is critical for the transformation of cybersecurity into a science. To better characterise what an ontology is, it is necessary to start from the notion of vocabulary.

Vocabularies define concepts and describe how they relate to each other. In a vocabulary, concepts are denoted by words, combined by complying with syntactic and semantic rules. In this regard, human communication is effective only insofar as the vocabularies used by the interlocutors—and therefore their conceptual models—are relatable. The less ambiguous a definition is, the more such a definition will be suitable to formal representation; minimising the ambiguity depends on the identification of the entities a concept denotes (extensional semantics), and on detailing the context of usage (intensional semantics). A thorough analysis of the distinction between intensional and extensional semantics is provided in (Kripke 1980).When such rigorous definitions are represented using a formal language, for example, one based on logical expressions or axioms, vocabularies become formal ontologies.

Moreover, when logical structures are encoded into a machine-processable language, such as RDF or OWL (http://www.w3.org/TR/owl2-rdf-based-semantics/), formal ontologies take a computational form, and can be used alongside other semantic technologies, such as search engines, automatic reasoners, and knowledge-management tools, among others. If shared vocabularies enable human communication, computational ontologies are crucial when human-computer interaction is required. Computational ontologies represent an important bridge between Knowledge Representation and Computational Lexical Semantics, and form a continuum with vocabularies; here 'lexicon' and 'vocabulary' are used as synonyms (Lenci, Calzolari & Zampolli 2002).

As observed by Mundie and McIntire (2013) and Dipert (2013), this strong dependence between vocabularies and ontologies is particularly evident in cybersecurity, the domain in which producing well-grounded natural language definitions (that are adopted, for instance, in military doctrines) is preparatory to the job of the ontology developer, which consists of providing formal representations of those definitions for computation in risk-management, intrusion prevention, and operational support systems, among others. Risk plays a prominent role in practical cybersecurity, and for this reason it is one of the most important conceptual structures that an ontology of cybersecurity needs to address.

A 2010 study sponsored by the DoD concluded that the two most important requirements for a science of cybersecurity are: (1) the construction of a common language and (2) the characterisation of a set of basic concepts, upon which to develop a shared understanding of the cybersecurity domain and design adequate security protocols for networked systems (MITRE 2010). But if XML-based languages, such as RDF and OWL have reached a mature stage of development, in compliance with requirement (1), the relatively low impact that ontologies have had in cybersecurity suggests that a shared understanding of basic domain concepts is yet to be achieved. The problem with fulfilling requirement (2) relies on the intrinsic complexity of the cybersecurity domain, a feature that clearly transpires from a recent report on quantification of cyber threats promoted by the World Economic Forum (http://www3.weforum.org/docs/WEFUSA_QuantificationofCyberThreats_Report2015.pdf); in this document experts pinpoint the bottleneck of cyber-threat assessment on the lack of "standardization and benchmarking of input variables", as conversely accomplished, they add, "by the car insurance industry." But if agreeing on the meaning of notions such as age and gender of drivers, weight and year built, claims history, and the like seems mostly straightforward, specifying the semantics of concepts such as system vulnerability, software usability, trust, and password strength, for example, might require advanced technical knowledge, fine-grained conceptual primitives, and expressive representation languages.

To satisfy these requirements, researchers should build rigorous conceptual models of cybersecurity, reusing and extending relevant ontologies when possible, and focusing on empirical evidence. Little effort has been put into this broad initiative so far, with exception of the MITRE project (Obrst, Chase & Markeloff 2012), recently reprised in Syed *et al.* (2015). This lack of a systematic approach could result in nothing but a heterogeneous variety of ontologies. Some examples are provided in the remainder of this section.

Fenz and Ekelhart (2009), for instance, propose an ontology based on four parts—security and dependability taxonomy, the underlying risk-analysis methodology, the concepts of the IT infrastructure domain, and a simulation-enabling enterprise—to analyse various policy scenarios. They carried out a comprehensive investigation, but their proposal is affected by an underspecified notion of risk, conceived as "the probability that a successful attack occurs". In fact, despite adequately conceptualising the 'risk space' as adversarial, they fail to account for the dependencies between attacks, system vulnerabilities, skills and level of expertise of the defenders, monetisation of information loss resulting from data breaches, and the like. An overly coarse and generic representation of cyber risk is a pervasive problem in the literature, as with the case of Assali, Lenne and Debray (2008) and Shepard *et al.* (2005), whose in-depth distinctions adopted to model cyberattacks are not matched by an analogous level of detail in defining cyber threats and in risk-assessment procedures.

By and large, the most popular modelling solution in risk-related ontology research seems to deal with the reification of risk assessment and threat quantification into the process of 'rating', whose attributes are expressed either qualitatively (by means of high, medium, and low dimensions in the Likert scale) or quantitatively (measuring the probability of a risk). The word 'reification' originates from Latin, and roughly translates as 'thing-making' (*res facere*). In this context, reification indicates the abstraction process according to which multiple related entities can be conceived as one. For instance, when one considers that the sky is blue, the common interpretation includes three entities, namely the 'sky' surface, the 'color' quality, and the specific value of it, such as 'blue'; by means of reification, those three entities can be represented as single thing, such as 'the blue coloration of the sky'. Temporal boundaries can be added to reification, so that more refined conceptual models are possible (for example, the orange coloration of the sky during sunset).

The theory of reification is widely used in the epistemology of mathematics (Sfard & Linchevski 1994). For instance, an algebraic equation can be seen as a unique object, or as a set of variables connected by operations such as subtraction, addition, and equality. In computational languages, reification is commonly adopted as a method to bypass expressivity limits. In the Resource Description Framework (RDF; https://www.w3.org/RDF/), for instance, a relation between *n* entities (with n>2) can be represented as a single statement about those entities, as in the example above. In this regard, one could represent that a given cyber vulnerability exposes a system to a certain risk factor (n=3) by asserting a risk-rating statement about the vulnerability of that system. For details on RDF reification, please consider: http://www.w3.org/TR/swbp-n-aryRelations/#RDFReification.

An alternative approach comes from Enterprise Risk Management (ERM), an area that concerns the identification, assessment, and mitigation of operational risk. For instance, Lykourentzou *et al.* (2011) focus on six subclasses of events: (1) failure, (2) infrastructure disruption, (3) occupational incident, (4) fraud, (5) disaster, and (6) attack, by binding each of these event types to a wide spectrum of 'root causes' and 'treatment plans' to address risk factors. ERM approaches can help not only to identify risk-related event patterns, but also to elicit the behavioural patterns in the adoption of risk-management practices.

These initial considerations suggest that one of the main problems affecting ontologies of cybersecurity is that cyber risk is ambiguously or too generically defined. In the next section, the authors discuss this problem more in depth. Another critical aspect, as mentioned above, is the lack of a systematic approach in ontology development: but how can a comprehensive procedure of ontology creation be put into effect? To this end, the researchers conducted an analysis of the state-of-the-art ontologies of cybersecurity centred on the notion of cyber risk. The methodology and the results of this analysis are described in the section below titled, Empirical Examination 2: Ontology Developers' Characterisation of Cyber Risk.

## Quantification of Cyber Risk

Many concepts, including cyber risk, cannot be defined in a pragmatic way without specifying an approach to quantifying them. Thus, no definition or ontology of cyber risk can be fully satisfactory unless one adds a way to quantify the risk. However, development of approaches to quantifying cyber risk has proven to be challenging.

For example, although a number of algorithms are used to quantify risk (for example, so-called risk-scoring algorithms), they remain limited to *ad hoc* heuristics, such as simple sums of vulnerability scores or counts of things such as missing patches or open ports. Weaknesses and the potentially misleading nature of such metrics have been pointed out by a number of authors, including Jansen (2009) and Bartol *et al.* (2009). It is important to consider that the individual vulnerability scores are dangerously reliant on subjective and human qualitative input, which is potentially inaccurate and expensive to obtain. Further, the total number of vulnerabilities may matter far less than how vulnerabilities are distributed over hosts or over time. The presence of vulnerabilities is often unknown and depends on characteristics of computer hosts in a complex manner (Gil, Kott & Barabási 2014). Similarly, neither the topology of the network nor the roles and dynamics of inter-host interactions are considered by simple sums of vulnerabilities or missing patches. In general, there is a pronounced lack of rigorous theory and models regarding how various factors might combine into quantitative characterisation of true risks, although there are initial efforts, such as Lippman *et al.* (2012), to formulate scientifically rigorous methods of calculating risks.

Cognitive biases are also evident in how humans deal with cyber risk. Researchers note that cyber defenders exhibit over-reliance on intuition; with few reliable statistics on cyberattacks, decision makers rely on their experience and intuition, both of which are fraught with cognitive biases. For example, when faced with the trade-off between a certain loss in the present (for example, investing in improved security) and a potential loss in the future (consequences of a cyber incident), a common risk-aversion bias is to avoid considering future adverse events and consequences. In a related observation, cyber defenders tend to believe that their particular organisation is less exposed to risks than other parties, particularly if they feel they have a degree of control over the situation; some refer to this as optimistic bias. Many are more afraid of risks if they are vividly described, easy to imagine, memorable, and if they have occurred in the recent past (related to what is called availability bias). Finally, a common bias is to ignore evidence that contradicts one's preconceived notions, such as the confirmation bias described by Julisch (2013). With respect to the optimistic bias mentioned above, it is important to note that individuals distinguish between two separate dimensions of risk judgment—the personal level and the societal level. Individuals display a strong optimistic bias about online privacy risks, judging themselves to be significantly less vulnerable than others to these risks. Internal belief (perceived controllability) and individual difference (prior experience) significantly modulate optimistic bias (Cho, Lee & Chung 2010). There is a tendency for individuals to interpret ambiguous information or uncertain situations in a self-serving direction. Perceived controllability and distance of comparison target influence this tendency (Rhee, Ryu, & Kim 2012).

Issues of cyber risk challenge communications and understanding between individuals involved in the use and defence of cyber assets (Nurse *et al.* 2011). This is exacerbated by barriers between individuals of different roles and backgrounds in cyber defence. For example, cyber experts see users both as potential cyber-defence resources, but also as sources of cyber risks—accidents and potential insider threats. Cyber defenders currently lack detailed knowledge of their users' information-security performance, which could help maintain trust relations for effective collaboration (Albrechtsen & Hovden 2009). In evaluating threats and prioritising how to respond to them, experts tend to use probability rather than consequences as a basis for evaluating risk. This may bias cyber defenders toward well-known threats instead of high-risk emerging threats. Finally, the adversarial nature of cyber defence, for example the critical role played by the intelligent adaptive adversary in influencing the quantification of cyber risk, is a major challenge. In criticising the approaches in which risk is quantified as a product of probability and consequence (impact) of an adverse event, Cox (2008) points out that such approaches force "practitioners to try to use and interpret numbers that have no clear conceptual definitions and that do not model the planning, learning, and adaptive replanning of intelligent attackers". The next section of this paper explores whether practitioners actually try to use such numbers or decline to do so—something Cox did not attempt to determine and which, to the knowledge of these researchers, has not been attempted before in a quantitative, empirical manner.

There a few examples of research that attempt to deal explicitly with adversarial aspects of cyber defence. For instance, Rios and Insua (2012) discuss how, in the framework called Adversarial Risk Analysis, a defender assesses the probabilities of the adversary's actions before computing the maximum expected utility defence. Rios and Insua point out the possibility of applying this approach, originally formulated in the context of counterterrorism, to a cybersecurity scenario, and note that cyber scenario differs by large and uncertain number of adversaries.

## Empirical Examination 1: Cyber Practitioners' Descriptions of Cyber Risk

The discussion now turns to an empirical study conducted by the authors that focused on the following research question: Do cybersecurity practitioners use a concept of cyber risk as reflected in prevailing definitions, that is, centred on likelihood and impact (or, probability and consequence) of an undesirable cyber event? The short answer is that they do not. This examination systematically investigated the structure and content of a large number of posts on a popular and well-regarded website frequented by cybersecurity practitioners: seclists.org. The posts selected contained descriptions of specific cyber risks, identified the common components of such descriptions, and calculated frequencies of various components and their co-occurrences.

## Data and analysis

The authors performed a systematic series of searches on the website seclists.org. The website contains a large number of continually updated posts whose authors appear to be mostly hands-on cybersecurity professionals. The posts typically inform the readers about vulnerabilities and other security issues, ask for related information, or reply to earlier posts. The search for the word 'risk' yielded a total of about 51,100 posts containing this term. The examination of these posts revealed that they fall into three categories.

The first category consisted of formal descriptions of vulnerabilities that require inclusion of the risk-severity assessment. These included standardised statements similar to the following example: 'The security risk of the persistent vulnerability is estimated as medium with a CVSS count of 4.1', without necessarily specifying what was seen as the risk. The authors did not analyse the posts in this category. The second category of posts included the term 'risk' as an adjective or qualifier or within a compound term, for example 'risk assessment', without actually describing a specific risk. The third category of posts (or replies to posts) contained a detailed description of a specific risk, often introduced by such words as 'The risk is that…' and similar phrases. After a preliminary manual inspection, scripts were developed to search the posts for several lexical constructs that were most likely to offer a description of a risk, for example, 'the risk is that…'. In these posts, it was clear from the content that the posts' authors intended to provide a description of what they believed to be risks and explicitly called such circumstances risks. The authors' analysis on this last category consisted of 34 posts authored from 2005-2014.

Each post was manually inspected to identify key semantic components of the risk description. In the most general case, a description included some of the following: the description or the system or network under consideration, the modification planned for the system, the vulnerability within the system, the method by which a threat actor might be able to exploit the vulnerability, the follow-on exploitation actions by the threat actor, the counteractions (mitigations or workarounds) available to the defenders. **Table 1**, below, shows the semantic components of risk descriptions in several representative cases. The authors also looked for any discussion of likelihood and impact associated with the risk and then tabulated the frequency of occurrences and co-occurrences of the description components.

|  | **Case #1** | **Case #2** | **Case #14** |
| --- | --- | --- | --- |
| **State of the system** | "Active Record with query API" | "snort station, couple sensor tap ports, MySQL DB, schema from Snorby" | "AirDroid v2.1.0 on CleanROM 8.1 Core Edition, phone Galaxy S3" |
| **Vulnerability** | "columns named identically to the table" | "remote sensors…no control of mothership DB server.." | "If you accept the connection request, browser will be granted full access to the phone" |
| **Threat action (exploit)** | "following code could return additional records: SecurityToken" | "threat steals DB authentication credentials from the .conf file" | "threat can gain access to the phone without the screen unlock" |
| **Threat follow-on** | N/A | "could delete large portion of snort central database" | N/A |
| **Mitigation** | "call to_s on the value from params…" | N/A | N/A |

**Table 1:** Components of a risk description in three representative cases

## Findings

The risk descriptions are most notable for what they do not include. They mention neither the impact of an undesirable event nor its extent. They also do not mention the likelihood of the event. In other words, recalling the quote from Cox (2008) from the previous section of this paper, the practitioners in fact do not "use and interpret numbers that have no clear conceptual definitions". This could lead to the conclusion that the practitioners do not use such numbers because the numbers do not model the planning, learning, and adaptive replanning of intelligent attackers.

A few partial exceptions have been noticed. In one case, the description inquired about the 'likeliness' that an exploit was available for a particular exposure. In another case, the author of a post argued that a particular vulnerability—a simultaneous occurrence of unusual factors—is very unlikely to happen in practice. Neither of these two exceptions refers directly to the likelihood of the final outcome, and neither attempted a quantitative estimate of the likelihood.

In several cases, the word 'impact' was used. It referred to creating a condition in which an attacker obtains an opportunity to execute follow-on exploitation action. In these cases, the word 'impact' did not refer to damage imposed on the operations of the system, and no attempt was made to assess the extent of this impact.

On the other hand, almost invariably, risk is defined by a description of a vulnerability (or a reference to a known vulnerability) followed by a description of one or more exploits that can take advantage of the vulnerability. This form of risk description is found in 78% of all cases. In addition, 8% of cases describe a vulnerability without explicitly mentioning an exploit, but clearly imply that an exploit exists or that a competent reader would be aware of an applicable exploit. In addition, 4% of cases focused on an exploit while merely implying a vulnerability. In total, the overwhelming majority of all cases described a specific risk by implicitly or explicitly specifying a pair vulnerability-exploit.

Unsurprisingly, virtually all cases also described a configuration (an initial state) of the system as a necessary background information. Also, 32% of the cases included a description of a suggested defensive action—a preventive or mitigating measure. And 21% included a description of follow-on actions that the attacker would be able to execute after a successful initial exploit.

Overall, it was found that cybersecurity practitioners describe cyber risk as a tuple (system configuration/state/operation, vulnerability, exploit, follow-on exploitation options, counteractions), where the last two elements are optional. Contrary to commonly cited definitions of cyber risk, concepts such as 'likelihood of an event' and 'the extent of its impact' play no discernible role.

In interpreting these observations, it is necessary to question whether the authors' observations were focused on the appropriate segment of the cybersecurity community. Perhaps cybersecurity professionals whose posts were analysed in this research were unfamiliar with high-level functions and missions of their employers, and preferred to focus on low-level details of their responsibilities—compliance, vulnerabilities, exploits, patching, and such—and therefore would not be in a position to discuss high-level, business-wide risks.

This was found to be unlikely for three reasons. First, if the practitioners represented on the seclists.com website are concerned and comfortable with likelihood and impact of low-level events, they would discuss such matters in describing risk. Yet the data show that they do not.

Second, although it is a limitation of this study that the demographics of the practitioners are unknown, it can be hypothesised that a high percentage of them work for small and medium enterprises (for example, a high fraction of respondent to surveys of cyber workforce noted that they work for small and medium enterprises) (SANS 2014; Suby 2013). In such businesses, it is common to have very few cyber experts (sometimes only one) in the entire enterprise. Such an expert often reports directly to the owner of the business or to a high-level executive. The expert has a broad view of the business and its functions and would be relied upon to report and to elucidate the business ramifications of cyber risks. If likelihood and magnitude of business impact were meaningful to him or her, he or she would likely mention them in his or her posts. Yet the data do no show this.

Third, if cybersecurity professionals whose posts were analysed in this research were unfamiliar with high-level functions and missions of their employers, then information security managers at higher level positions within a corporation would be expected to have a more comprehensive view of such matters. Correspondingly, they would display a more profound understanding of cyber risk. However, based on interviews with multiple Chief Information Security Officers (CISOs), Libicki, Ablon & Webb (2015) conclude that these executive officers show little ability to conceptualise and quantify cyber risks. As they note,

> CISOs, vendors, and the security community as a whole focus on threats, rather than risks. The ability to understand and articulate an organization's risk arising from network penetrations in a standard and consistent matter does not exist, and will not for a long time.

For these reasons, the authors conclude that the segment of cybersecurity professionals whose descriptions of cyber risks were studied here is appropriate for this research.

## Empirical Examination 2: Ontology Developers' Characterisation of Cyber Risk

While the authors' first empirical study focused on the perspective of cyber practitioners, in the second study—the subject of this section—the focus is on the perspective of a very different group of professionals within the cyber defence community, the developers of ontologies. Unlike cyber practitioners, ontology developers are explicitly tasked with in-depth analysis and interpretation of concepts related to cyber risk. Therefore, similarities and differences in results of two empirical studies are significant and instructive. This second empirical study aimed to answer two interrelated questions: (1) How many risk-related concepts are covered by the state of the art on ontologies of cybersecurity? and (2) To what extent do existing ontologies of cyber risk address the adversarial dimension?

## Data and analysis

The authors took into consideration 10 ontologies of cybersecurity the names and sources of which are listed in the caption to Table 2. This selection was driven by availability of appropriate technical documentation (articles, web repositories, and technical reports). In order to compare the ontologies, concept density (Prévot, Borgo & Oltramari 2005) was adopted, a reference measure from 0 to 1 denoting the proportion between the number of concepts covered by an ontology and the number of terms of a vocabulary of reference. A centralised vocabulary of cyber risk and cybersecurity terms was not available. To bypass this obstacle, the authors created a vocabulary from different resources: (Avižienis et al. 2004); (Karyda et al. 2006) (Donner 2003); (Obrst, 2012); (Lin, 2012); (Fenz, 2009); (Schumacher, 2003); (Souag et al. 2012); (Deloitte, 2015); (MITRE, 2011). The resulting list of 45 terms is shown in the first column of Table 2. The resulting vocabulary contains only a few adversarial notions, namely cyber threat, cyberattack, and intention; this initially shows that the terminology the authors were able to find is still inadequate to suit a holistic approach to cybersecurity.

For reasons of readability, **Table 2**, above, does not include the vocabulary definitions from the examined documentation. Some examples follow. Avižienis *et al.* (2004) define 'failure' as "an event that occurs when the delivered service deviates from correct service"; Fenz and Ekelhart (2009) define 'cyber vulnerability' as "the absence of a proper safeguard that could be exploited by a threat". Bodeau and Graubart (2011) define a 'cyber attack' as

> an attack on cyber resources. The attack is typically, but not necessarily, carried out by cyber means. The attack may be intended to adversely affect the cyber resources, or to adversely affect the missions, business functions, organizations, or populations that depend on those resources.

However, the specificity of a domain guarantees substantial convergence of term usage across different ontologies, as indicated by state of the art in information retrieval (Hersh 2008). For example, cybersecurity practitioners regularly talk about 'detection' and 'origin/source' without necessitating any disambiguation.

| Common Vocabulary for Risk M | CESO | ONTIDS | CRATELO | OMG | M4D4 | CORESEC | NRL | WALI | CycSecure |
|---|---|---|---|---|---|---|---|---|---|
| Alert | | X | | | X | | | X | |
| Asset ( ⊂ tangible, untangible) | X | | X | X | | X | X | | X |
| Benefit | | | | X | | | | | |
| Configuration | | | X | X | X | | | | |
| Consequence | X | | X | X | | X | | | X |
| Control | | | | X | | | | | |
| Cost | | | X | X | | | | | |
| Countermeasure | | | X | X | | | | X | X |
| Credential | | | | X | | X | | | |
| Cyber attack | X | X | X | X | X | X | | X | X |
| Cyber defense | | | X | X | | | | | |
| Cyber exploitation | | X | X | X | X | | | | |
| Cyber incident | | | | X | | X | | | |
| Cyber operation | | | X | X | | | | | |
| Cyber response | | | X | X | | | | | |
| Cyber risk | | | X | X | | | | | |
| Cyber threat | | | X | X | | | | | |
| Cyber vulnerability | | X | X | X | X | X | | | X |
| Dependability ( ⊂ attributes) | | | | X | X | | | | |
| Detection | X | X | X | X | X | | | | |
| Fault | | | X | X | | | | | X |
| Failure | | | X | X | | | | | X |
| Impact | | | X | X | X | X | | | |
| Intent | | | X | X | | | | | |
| Likelihood | | | X | X | | X | | | |
| Mission | | | X | X | | | X | | |
| Network | | X | | X | X | | | | |
| Origin/Source | X | X | X | X | X | | | | |
| Payload | | | X | X | | | | | |
| Report | | | | X | X | | | | |
| Risk | | | X | X | X | | | | |
| Risk assessment | X | X | | X | | | | | |
| Risk factor | | | X | X | | | | X | |
| Risk identification | | | | X | | | | | |
| Risk metric | | | X | X | | X | | | |
| Risk mitigation | | | | X | | | | | |
| Risk monitoring | | | | X | | | | | |
| Security/Risk policy | | | | X | X | | | X | |
| Security protocol | | | | X | | | | | |
| Situation | | | X | X | | | | | X |
| Service | | X | X | X | X | | | X | |
| Stakeholder | | | X | X | | | | | |
| Target | X | X | X | X | X | | | | X |
| Threat | | X | X | X | | X | | | |
| Treatment | | | | X | | | | | |

**Table 2**: Vocabulary concepts covered by the analysed ontologies: **CESO** (Ormrod, Turnbull, & O'Sullivan 2015); **ONTIDS** (Sadighian *et al*. 2014); **CRATELO** (Oltramari *et al*. 2014); **OMG** (http://www.threatrisk.org); **M4D4** (Morin *et al*. 2009), **CORESEC** (de Azevedo *et al*. 2010); **NRL** (Kim, Luo & Kang 2005); **WALI** (Wali, Chun & Geller 2013); **CycSecure** (Shepard *et al*. 2005); **ORM** (Lykourentzou *et al*. 2011)
0

Together with concept density, Prévot, Burgo, and Oltramari (2005) also define constraint density, which assesses the types of formal constraints used in an ontology. In general, a formal constraint is a restriction that applies to properties of entities. For instance, a definition can state that every person has 'only one' Social Security Number (SSN). If this cardinality restriction is violated, one might infer that there was a problem in the computer system that generates the SSN, or that there was an identity theft. Since all ontologies but CycSecure were implemented in OWL-DL, their formal constraints share the same expressivity, namely that of Description Logics (Borgida 1996): union or disjunction of concepts, negation, universal and existential restriction of concepts, and the like (Grigoris & Van Harmelen 2004). For instance, CRATELO (Oltramari *et al.* 2014) defines a TCP network session as being constituted, at minimum, by a sequence of two packets (the SYN packet, transmitted from the source to the destination node, and the ACK packet, the acknowledgement that the destination node has received the SYN packet). Here the formal constraint of 'cardinality restriction' (minimum 2) is applied to the OWL object property 'has_member' with domain 'TCP Network Session' and range 'packet'. ONTIDS uses a similar semantic construct to restrict the number of 'stages' in a cyber attack, for example, a minimum of one. CycSecure, on the contrary, uses a second-order language (Lenat 1995) that allows for quantification over relations; for instance, it could be said that some relations of the type 'targeting' hold between cyber entities (for example DDoS – targets – website), some hold between physical entities (for example card trapping – targets – ATMs), while others hold between cyber and physical entities (for example Stuxnet – targets – nuclear plants), and so on. Although CycSecure is more expressive than the OWL-DL ontologies, this greater expressivity affects its interoperability with existing resources and systems. CycSecure's being limited by proprietorship also affects interoperability and reusability of the resource.

## Findings

This section summarises the most interesting findings of the examination. **Figure 1**, below, shows that OMG and CRATELO have the highest 'concept density', respectively 0.97 and 0.66. This result emerges from the systematic approach behind the design of both ontologies, whose models encompass foundational, middle, and domain levels of conceptualisation; and both use some version of DOLCE (Masolo et al. 2002) as foundational layer. OMG is more mature than CRATELO and more effectively grounded on practical cybersecurity, an important factor that contributes to the high score.

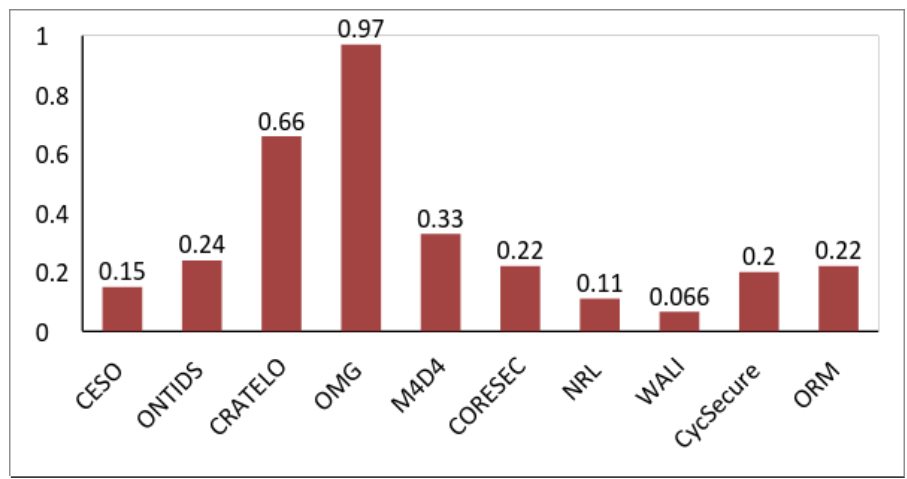

**Figure 1:** Concept density of the analysed ontologies

**Figure 2**, below, represents the frequency of terms across the analysed ontologies.

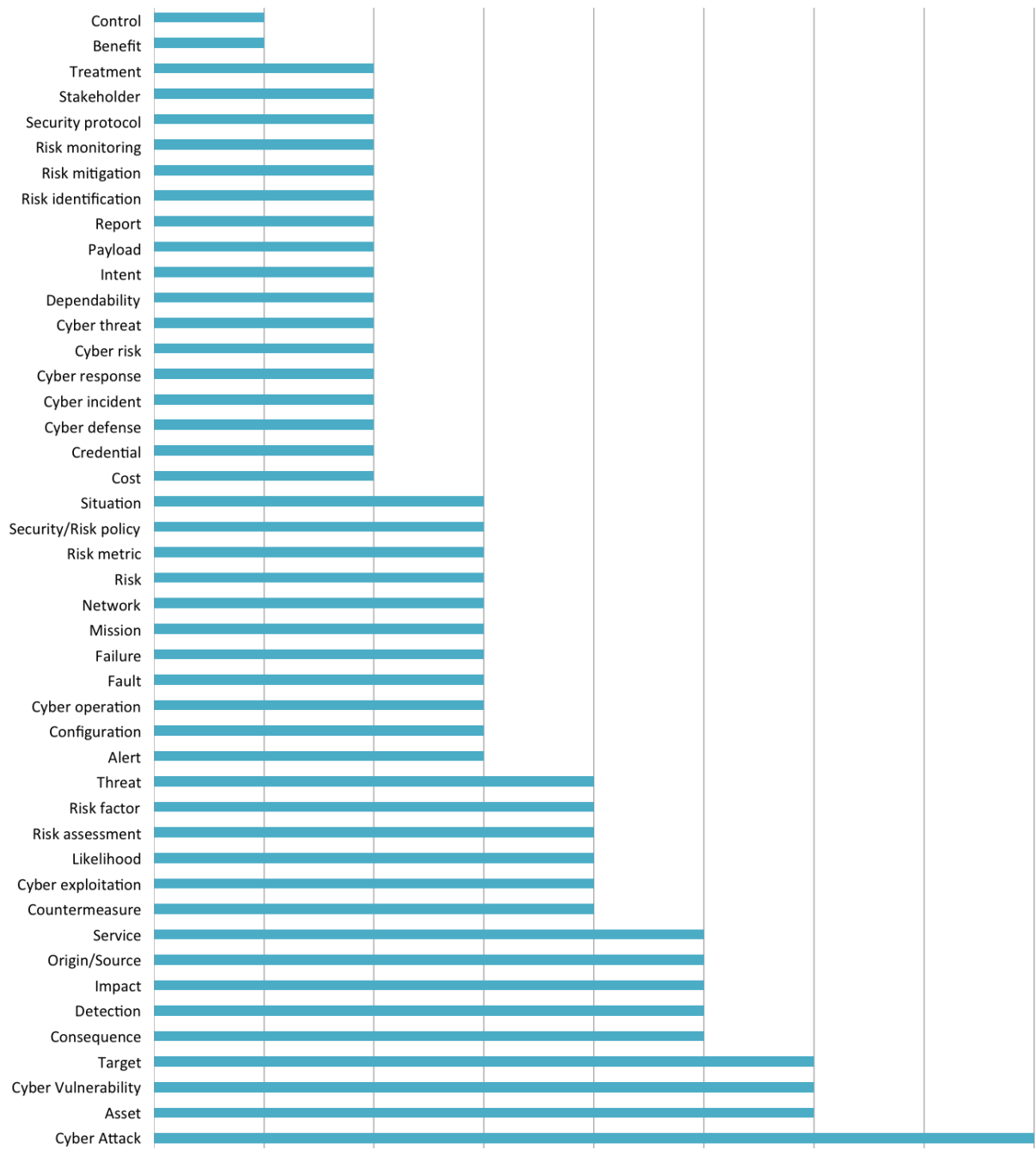

**Figure 2:** Frequency of occurrence of terms across the considered ontologies: the frequency of 'Control' is two, i.e., it appears in only two ontologies; the frequency of 'Cyber Attack' is eight, i.e., it appears in eight ontologies.

Only nine terms out of 45 occur five or more times; within these nine terms, 'cyberattack' is the most frequent, with eight occurrences, followed by 'asset', 'cyber vulnerability', 'target' (6 times), 'service', 'origin', 'impact', 'detection', and 'consequence' (5 times). Most of the ontologies have sparse coverage of the key notions included in the vocabulary. 'Cyber attack', 'vulnerability',

'asset', 'origin', and 'target' are among the most recurring concepts in the selected ontologies. This indicates a vulnerability-centric view of risk, focused on the asset(s) targeted by an attack. The 'likelihood' of an undesirable event is used by only three ontologies, indicating a marginal role of this concept (an outcome of Empirical Examination 1 also).

Traditional dimensions of risk seem to play a negligible function in the ontologies of cybersecurity that were evaluated. For instance, the 'impact' of an attack generally takes only three possible values, namely 'low', 'medium', or 'high' in the ontologies that incorporate this concept (CRATELO, OMG, M4D4, CORESEC, and ORM).

Instead, there is a notable tendency to perceive cyber risk in adversarial terms (by using 'attack', 'vulnerability', 'target', and similar words), as referring to an interaction of defenders with attackers. In this respect, a redefinition of cyber risk needs to be centred on the possibility of exploit of vulnerabilities, which is the *condition sine qua non* for an attacker to succeed in forcing a computer system from a healthy to a dangerous state.

These considerations argue for a new formal treatment of cyber risk that should include rigorous logical and mathematical models aligned with common sense understanding of cyber risk by practitioners. Examination 2 shows that only two ontologies of cybersecurity currently seem to be fit for the task, namely OMG and CRATELO; M4D4 and CORESEC do represent the adversarial element, but their 'concept density' is too low to constitute a satisfactory account of the domain. CRATELO could benefit from a systematic grounding on the domain knowledge of cybersecurity practitioners, a signature characteristic of OMG's high coverage.

## Discussion and Conclusions

This paper has explored the conceptualisation of cyber risks as reflected in utterances and products of two groups of professionals: cybersecurity practitioners and creators of ontologies pertaining to cybersecurity. Both groups stress the adversarial nature of risk in the cybersecurity domain. In the case of practitioners, they discuss cybersecurity risk primarily with respect to an intelligent adversary who would likely exploit a vulnerability. In case of ontology developers, they place greater emphasis on concepts like attack, vulnerability, and target, also reflective of focus on an adversary. Both groups perceive the concept of vulnerability, along with its exploitation by an attacker, as exceptionally salient.

At the same time, neither practitioners nor ontologists pay comparable attention to the concepts traditionally associated with risk, such as probability or likelihood of an adverse event, and the cost of consequences or impact of the event. Such concepts, which are canonical in most definitions inspired by traditional definitions of risk, are mentioned very infrequently in discourses of practitioners and with only moderate frequency by ontologists. In fact, only three out of 10 ontologies include both 'likelihood' and 'consequences' (or 'impact'). Even then, it is unclear whether these are intended to be quantified.

What could explain this apparent reluctance of these two groups of professionals to embrace the traditional conceptualisation of risk in application to cyber conflict? It is beyond the scope of this paper to attempt a rigorous proof—assuming such a proof is possible—that these attitudes of cybersecurity professionals are determined by a specific set of causes. However, arguments for rejecting traditional risk concepts, including the concepts' reliance on assessments of probability and consequence, have been discussed in at least one other adversarial domain, terrorism (Cox 2008).

Similar arguments apply to the cyber domain. As in the case of assessing the risk of a terrorist act, adaptive planning of an intelligent adversary makes estimating probability and consequence exceedingly difficult and potentially misleading. The defender's estimate of probability of an attack, argues Cox (2008), is largely meaningless because the probability would be greatly affected by the attacker's assessment of the defender's estimate; if the attacker perceives the defender as highly vigilant, the attacker will focus on another target, thereby negating the defender's assessment of the attack probability. Likewise, the defender's assessment of probability of the attack's success is impossible without detailed considerations of strategies and interactions between attacker and defender. Finally, the defender's assessment of the successful attack's consequences is highly questionable, as it depends on knowledge of the attacker's intent and specific actions, as well as on perceptions and reactions of relevant third parties.

These ambiguities explain why cyber practitioners rarely refer to the quantity of an impact or the probability of an adverse event when describing cyber risk. This also explains why ontologists, who are likely to construct their ontologies primarily from discussions with cyber professionals, also place little emphasis on traditional components of risk–probability and consequence. The authors do not presume that cybersecurity professionals do not value the importance of knowing traditional measures of likelihood and impact of event. Rather, the empirical evidence suggests, just as Cox (2008) does, that the professionals perceive these numbers as exceedingly ambiguous and impractical to determine. To the best of the authors' knowledge, while theoretical arguments for this rejection of traditional concepts of cyber risks have existed for at least a decade, this research offers the first quantitative empirical evidence that such a rejection by cybersecurity professionals is indeed observed in real-world practice.

If one must conclude that the traditional conceptualisation of risk is inapplicable to cyber domain, then what would replace it? A detailed proposal and its justification as a topic of future research goes beyond the scope of this paper, but it is appropriate here to suggest a possible approach. The authors envision an approach as described by Colbert, Sullivan and Kott (2017) that fully embraces the adversarial nature of cyber defence and quantifies cyber risk via a systematic and comprehensive evaluation of adaptive interactions between attacker and defender.

To this end, useful lessons could be learned from military wargaming. Indeed, cyber risk belongs firmly to the realm of adversarial decision-making and has little meaning outside of a process geared toward decisions made under threat of adversarial actions' pre-empting and counteracting those of the defenders. As such, practical insights evolved in the long history of military decision-making are highly relevant to cyber risks. For example, the U.S. Army's formal Military Decision Making Process (US Army 2012; Kott *et al.* 2002; Kott & Corpac 2007) is a rich source of practical techniques, particularly wargaming, which could be adapted and partly automated by cyber practitioners facing decision-making in the presence of sophisticated cyber adversaries.

In military practice, wargaming products typically include sets of interrelated actions of the defender and the attacker, the resources applied, the losses experienced by both sides, conditions under which different strategies may become relevant, and alternative outcomes that would emerge depending on observations and choices made by the opponents. It is this rich collection of multi-dimensional information that is needed to formulate a meaningful characterisation of risk in cyber confrontations as well.